\documentclass[12pt]{article}
\usepackage{a4wide}
\usepackage{epsf}  

\def\be{\begin{equation}}
\def\ee{\end{equation}}
\def\bea{\begin{eqnarray}}
\def\eea{\end{eqnarray}} 
\def\nn{\nonumber \\}

\def\Ortin{Ort{\'\i}n }

\def\part{\partial}
\def\tfrac#1#2{{\textstyle{#1\over #2}}}

\def\hR{{\hat R}}

\def\tR{{\tilde R}}

\def\hg{{\hat g}}

\def\tg{{\tilde g}}

\def\da{{\dot a}}
\def\dda{{\ddot a}}

\def\tLambda{{\tilde{\Lambda}}}

\def\hsqrtg{\sqrt{|{\hat g}|}}

\def\makeatletter{\catcode`\@=11}
\makeatletter
\def\mathbox#1{\hbox{$\m@th#1$}}%
\def\math@ccstyles#1#2#3#4#5#6#7{{\leavevmode
      \setbox0\mathbox{#6#7}%
      \setbox2\mathbox{#4#5}%
      \dimen@ #3%
      \baselineskip\z@\lineskiplimit#1\lineskip\z@
      \vbox{\ialign{##\crcr
             \hfil \kern #2\box2 \hfil\crcr
             \noalign{\kern\dimen@}%
             \hfil\box0\hfil\crcr}}}}
\def\mathaccstyles{\math@ccstyles\maxdimen}
\def\maththroughstyles{\math@ccstyles{-\maxdimen}}
\def\unitmatrixDT%
 {\maththroughstyles{.45\ht0}\z@\displaystyle {\mathchar"006C}\displaystyle 1}
\begin{document}

\rightline{IFT-UAM/CSIC-00-19}
\rightline{hep-th/0005116}
\rightline{\today}
\vspace{1truecm}

\centerline{\Large \bf Curved dilatonic brane-worlds and} 
\vspace{.5cm}

\centerline{\Large \bf  the cosmological constant problem}
\vspace{1truecm}

\centerline{
    {\bf Natxo Alonso-Alberca}\footnote{E-mail address: 
                                  {\tt natxo.alonso@uam.es}}, 
    {\bf Bert Janssen}\footnote{E-mail address: 
                                  {\tt bert.janssen@uam.es}}
    {\bf and} 
    {\bf Pedro J. Silva}\footnote{E-mail address: 
                                  {\tt psilva@delta.ft.uam.es}}}
  
\vspace{.4truecm}
\centerline{{\it Instituto de F{\'\i}sica Te{\'o}rica, C-XVI,}}
\centerline{{\it Universidad Aut{\'o}noma de Madrid}}
\centerline{{\it E-28006 Madrid, Spain}}
\vspace{2truecm}

\centerline{\bf ABSTRACT}
\vspace{.5truecm}

\noindent
We construct a model for dilatonic brane worlds with constant curvature on the 
brane, i.e. a non-zero four-dimensional cosmological constant, given in 
function of the dilaton coupling and the cosmological constant of the bulk.
We compare this family of solutions to other known dilatonic domain wall 
solutions and apply a self-tuning mechanism to check the stability of our 
solutions under quantum fluctuations living on the brane.

\newpage

\noindent
Recently the idea of brane worlds has received a lot of attention. In this 
picture our space-time has one (or more) extra non-compact spacial dimensions. 
In the scenario of Randall and Sundrum \cite{RS, RS2}, a three-brane in a
five-dimensional AdS-space was constructed, confining a Standard Model-like
gauge theory on the brane. Due to the warped form of the metric of the 
solution, a different approach to the hierarchy problem was presented. But it 
also turns out that in this picture, perturbations of the metric on the brane 
have a five-dimensional profile which is normalisable and localised  to the 
brane as if it were four-dimensional. Hence, an observer living in this brane 
world would see both gauge as gravity physics in the same way as an observer
in a four-dimensional space-time.

\noindent
Different generalisations of this picture including a dilaton field were given
in \cite{Youm}-\cite{KSS2} while on the other hand, generalizations to non-flat
brane worlds (without dilaton) appeared in \cite{AMO}, where a four-dimensional
cosmological constant was introduced via constant curvature branes. Here it 
was shown that, in the brane world scenario, the four-dimensional cosmological
constant is a geometrical property of the three-brane (namely its internal 
curvature), which in principle can be independent of the five-dimensional one.

\noindent
This shed new light on an old problem, namely why the observed cosmological 
constant in our universe is so small. One would expect that the non-zero vacuum
expectation value of Standard Model fields would generate a non-zero vacuum 
energy, which would result in an effective non-zero cosmological 
constant.\footnote{For a more general discussion on the cosmological constant 
problem, see \cite{Witt} and references therein.} Now 
that it turns out that in the brane world picture the four-dimensional 
cosmological constant is a geometrical property of the brane, one should ask
how this property is affected by quantum fluctuations of the field theory 
living on this brane. 

\noindent
In \cite{KSS, DR, KSS2} it was observed that fluctuations of the brane tension,
due to quantum corrections of the field theory living on a (flat) dilatonic 
brane, do not generate a four-dimensional cosmological constant. Via a 
self-tuning mechanism, the fluctuations in the brane tension can be absorbed 
into shifts of the dilaton, such that the quantum corrections do not curve the 
brane itself or the extra dimension. Thus, starting with flat brane worlds, no 
extra curvature, and hence no four-dimensional cosmological constant, is 
generated.

\noindent
Recent astronomical observations, however, point in the direction of a 
small positive, but non-zero cosmological constant. It is therefore 
interesting to look at brane-world models of non-zero curvature and see if 
there is a self-tuning mechanism working in these cases, which could protect
this curvature from quantum corrections. The aim of this letter is to 
combine the results of \cite{Youm}-\cite{KSS2} and \cite{AMO}, constructing 
dilatonic brane-worlds with a non-zero four-dimensional cosmological constant
and test the self-tuning mechanism on these solutions. The organisation of 
this letter is as follows: we start with the construction of the curved 
dilatonic brane-world solutions and then give a small discussion of these 
solutions, comparing the obtained solutions with a class of domain wall 
solutions given in \cite{CR}. Finally we will analyse the self-tuning mechanism
in the case of curved dilatonic branes.  

\vspace{.4cm}

\noindent
As a starting point let us consider the action of five-dimensional dilatonic 
gravity coupled to a brane source in the presence of a (five-dimensional) 
cosmological constant:
\bea
S&=&\frac{1}{\kappa}\int d^4 x \ dy  \ \hsqrtg 
                      \ \Bigl[ \hR + \tfrac{4}{3} (\part \phi)^2 
                                 - e^{\frac{\alpha}{3}\phi} \Lambda \Bigr] \nn
&& \hspace{1.3cm} - \int d^4 x \ \sqrt{|\bar g|}\  
                    e^{\tfrac{\beta}{3}\phi} \ V_0 \ , 
\label{action}
\eea
where $V_0$ is the tension of the brane source and 
$\bar{g}_{mn}=\hg_{\mu\nu} \delta^\mu_m\delta^\nu_n$ the induced metric on 
the brane. To solve the equations of motion,
we propose the following curved brane-world Ansatz:
\bea
ds^2 &=& a(y)^2 \tg_{mn} dx^m dx^n - dy^2 \nn
\phi &=& q \log a(y) + \phi_0    \ ,
\label{ansatz}
\eea
with $q$ and $\phi_0$  arbitrary constants.
The internal brane metric $\tg_{mn}$ depends only on the internal 
coordinates $x^m$. Plugging this Ansatz in the equations of motion of the 
action (\ref{action}) gives the following set of differential equations:
\bea
&& \tR_{mn} - \tg_{mn} \Bigr[ a\ \dda + 3\ \da^2 
           + \tfrac{1}{3}\ a^{\frac{\alpha q}{3} +2} 
                       e^{\frac{\alpha}{3}\phi_0} \Lambda \Bigr ] =
    +\tfrac{1}{6} \kappa V_0 \ a^{\frac{\beta q}{3} +2} \
                 e^{\frac{\beta}{3}\phi_0} \tg_{mn} \delta(y)   \ ,\nn
&& 4 \ a^{-1} \dda +\tfrac{4}{3}\ q^2 a^{-2} \da^2 
                       + \tfrac{1}{3}\ a^{\frac{\alpha q}{3}}
                           e^{\frac{\alpha}{3}\phi_0} \Lambda =
               -\tfrac{2}{3} \kappa V_0 \ a^{\frac{\beta q}{3}}   \
                    e^{\frac{\beta}{3}\phi_0} \delta(y)     \ , 
\label{eqns} \\
&& q \ a^{-1} \dda + 3 \ q \ a^{-2} \da^2 
            - \tfrac{\alpha}{8} \ a^{\frac{\alpha q}{3}}  
                   e^{\frac{\alpha}{3}\phi_0}\Lambda 
         = + \tfrac{\beta}{8} \kappa V_0\  a^\frac{\beta q}{3}  \ 
                e^{\frac{\beta}{3}\phi_0} \delta(y) \ ,
              \nonumber
\eea
where $\tR_{mn}$ is the Ricci tensor of the internal metric $\tg_{mn}$
and a dot denotes derivative with respect to $y$.
Analogous as in  \cite{AMO}, the first equation of (\ref{eqns}) can only be 
satisfied if the $y$-dependence vanishes: 
\be
a\ \dda + 3\ \da^2  + \tfrac{1}{3}\ a^{\frac{\alpha q}{3} +2} 
            e^{\frac{\alpha}{3}\phi_0}\Lambda 
+\tfrac{1}{6} \kappa V_0 \ a^{\frac{\beta q}{3} +2} \
                 e^{\frac{\beta}{3}\phi_0}  \delta(y) 
                     = \tLambda \ ,
\label{tLambda}
\ee
with $\tLambda$ an abritrary integration constant which we interpret as the 
four-dimensional cosmological constant in the  brane-world, since the first 
equation of (\ref{eqns}) now translates into $\tR_{mn} = \tLambda \ \tg_{mn}$.
The solution to the equations (\ref{eqns})-(\ref{tLambda}) is given by
\bea
ds^2 &=& \Bigl[ 1 - \tfrac{\alpha}{12}\
              \sqrt{- \Lambda } \ |y| \Bigr]^2 \
                   e^{\frac{\alpha}{3}\phi_0}\ \tg_{mn} dx^m dx^n - dy^2 \ ,\nn
\phi &=& -{\frac{6}{\alpha}}\ \log \Bigl[ 1 
         - \tfrac{\alpha}{12}\ \sqrt{- \Lambda} \ |y| 
             \Bigr]   \ ,
\label{holografic}
\eea
where the coordinate $y$ runs from 0 to $\tfrac{12}{\alpha{\sqrt{-\Lambda}}}$ 
and  the four-dimensional metric and 
cosmological constant satisfy
\be
\tR_{mn} = \tLambda \ \tg_{mn} \ ,  \hspace {2cm}
\tLambda = \tfrac{16 - \alpha^2}{48} \Lambda \ e^{\frac{\alpha}{3}\phi_0} \ .
\label{eq:lambdas}
\ee 
In the conformal frame, the above solution takes the form
\bea
ds^2 &=& \exp \Bigl( \tfrac{\alpha}{6}\ 
            \sqrt{- \Lambda}\ |z| \Bigr) 
                 \Bigl[ e^{\frac{\alpha}{3}\phi_0}\tg_{mn} dx^m dx^n - dz^2 \Bigr] \ ,\nn
\phi &=& -\tfrac{\sqrt{-\Lambda}}{2}\ |z|\ .
\label{conformal}
\eea
The brane tension and dilaton coupling in the source term are given in terms 
of $\alpha$ and $\Lambda$ by the jump equations:
\be
V_0 = \frac{\alpha}{2\kappa} \sqrt{-\Lambda} \ ,
\hspace{2cm}
\beta = \frac{8}{\alpha} \ .
\label{jump}
\ee
We thus see that the brane world are surfaces of positive (dS) or negative 
(AdS) constant
curvature, depending on the dilaton coupling $\alpha$.  It is remarkable that 
in the limit $\tLambda \rightarrow 0$ ($\Lambda\neq 0$), we do not recover the 
general ``flat'' dilatonic RS theory \cite{Youm, KSS, GJS}, but only a 
particular case $\alpha=\pm4$.\footnote{In \cite{GJS} the case $\alpha=4$ was 
erroneously identified with the case $\Lambda =0$ due to a coordinate 
singularity in the conformal frame. The holographic frame solution however is 
completely regular and can be seen as the limit of (\ref{holografic}) for 
$\tLambda =0$.} 
Therefore, in contrast to ``flat'' dilatonic RS theory, it is no longer 
possible to make contact with the original RS scenario \cite{RS, RS2}. 
Nor is it possible, in the limit $\alpha \rightarrow 0$ 
($\phi$ trivial), to make contact with the solutions of \cite{AMO}, as 
can clearly seen in the conformal frame (\ref{conformal}).
The solution for  $\Lambda = 0$ ($\tLambda\neq 0$) was given in 
\cite{KSS2}.
\vspace{.4cm}

\noindent
It is interesting to compare the solution (\ref{holografic}) with the 
dilatonic domain wall solutions found in \cite{CR}.  It is clear that 
(\ref{holografic}) can not be identified as one of the solutions given in 
\cite{CR}. This is due to the fact that the Ansatz used in \cite{CR} is 
different as our Ansatz (\ref{ansatz}). The main difference lays in the fact 
that the Ansatz of \cite{CR} considers branes with constant spatial curvature,
while the Ansatz (\ref{ansatz}) constructs branes with constant world volume 
(space-time) curvature. Still, there exists a particular solution that belongs 
to both classes, i.e. that can be obtained from each class as a special limit. 
For this case, the domain wall has to be flat, which implies 
$\alpha = \pm 4$ in equation (\ref{holografic}) and $M=k=0$ in the Type II 
solutions of \cite{CR}.
Furthermore, in the Ansatz of \cite{CR} the metric components 
$g_{tt}$ and $g_{ij}$ (the radial function in front of the spatial part of the 
brane world volume) should be identified. This particular solution is (in our 
notation, for $\alpha = 4$ e.g.):
\bea
ds^2 &=& \Bigl( 1 - \tfrac{1}{3}\sqrt{-\Lambda}\ |y|  \Bigr)^2 
               e^{\frac{4}{3}\phi_0}
             \Bigl[dt^2 - dx_i^2 \Bigr] - dy^2 \ , \nn
e^\phi &=& \Bigl( 1 - \tfrac{1}{3}\sqrt{-\Lambda}\ |y|  \Bigr)^{-3/2} \ .
\label{particular}
\eea 
An analysis, similar as the one done in \cite{CR} reveals that this solution 
is a static domain wall, due to the fact that for this case the potential is 
constant. This particular solution (\ref{particular}) was first given in 
\cite{LPSS}.

\vspace{.4cm}

\noindent
A straightforward analysis of the perturbations of the metric 
(\ref{holografic}) gives the following profile for the four-dimensional 
graviton
\begin{equation}
\psi (y) = \left[ 1-\tfrac{\alpha}{12}\sqrt{-\Lambda}\ |y| \right]^{2}\, ,
\end{equation}
which is normalizable in the interval 
$[0, \frac{12}{\alpha\sqrt{-\Lambda}}]$ 
and localized around $|y|=0$. Note that the dependence on dilaton coupling 
$\alpha$ is much weaker than in the ``flat'' dilatonic RS scenario, where we
had an exponential dependence on the coupling.

\noindent
Finally, note that, similar to other dilatonic brane world models, the 
solution (\ref{holografic})-(\ref{jump}) has a time-like naked singularity in 
$y=\frac{12}{\alpha\sqrt{-\Lambda}}$ ($z=\infty$). 

\vspace{.4cm}
\noindent
In the brane world picture, the four-dimensional cosmological constant 
$\tLambda$ is, rather than a input parameter of the Lagrangian, a geometrical 
quantity, related to the curvature of the domain wall. In \cite{KSS, DR, KSS2}
it was argued that for flat domain walls (i.e. $\tLambda = 0$), the quantum 
fluctuations of the gauge theory living on the brane can be absorbed in shifts
of the dilaton, which turns out to be a symmetry of the solutions. This 
mechanism, called self-tuning, makes that no extra curvature is generated and 
that the four-dimensional cosmological constant remains zero, even after 
quantum corrections. 

\noindent
Since recent astronomical observations seem to indicate that the cosmological 
constant might be very small, but non-zero and positive, a natural question to 
ask is how far this self-tuning mechanism can be extended, in particular to 
branes with non-zero curvature (i.e. solutions with non-zero four-dimensional 
cosmological constant). It turns out that no self-tuning mechanism is possible 
for the solutions we have constructed, due to the fact that the jump equation 
(\ref{jump}) does not depend explicitly on $\phi_{0}$. In fact, $V_{0}$ only 
depends on the bulk parameters and an arbitrary integration constant (which, 
without lost of generality, we set equal to 
$\frac{12}{\alpha\sqrt{-\Lambda}}$), indicating the position of the naked 
singularity. Changes of $V_{0}$ due to quantum fluctuations could only be 
absorbed by this integration constant, but this would mean that the positions 
of the singularities should change, which seems to have no sense. 
This seems to indicate that the self-tuning mechanism of \cite{KSS, DR, KSS2}
is only valid for the case of flat branes and does not provide a satisfying 
explanation of the cosmological constant problem in the light of the recent
astronomical observations.  

\vspace{.4cm}

\noindent
The family of solutions we have constructed does not include the $\alpha=0$ 
case, as can be seen from (\ref{holografic}) and (\ref{jump}).
With the Ansatz $\phi=\phi(y)$, the equations of motion for $\alpha =0$
read:
\begin{equation}
\label{eq:EOMa=0}
\begin{array}{rcl}
&& \ddot{\phi} + 4\, a^{-1}\dot{a}\dot{\phi} = 0 ,\\ 
&& a^{-1}\ddot{a} +  \tfrac{1}{3}\, \dot{\phi}^{2}
   +  \tfrac{1}{12}\Lambda = 0 ,\\ 
&& a^{-1}\ddot{a} + 3\, a^{-2}\dot{a}^{2} + \tfrac{1}{3}\Lambda
   = \tilde{\Lambda} a^{-2} . \,
\end{array}
\end{equation}
The first of these equations gives
\begin{equation}
\label{eq:dilgamma}
 \dot{\phi} = \gamma a^4\, ,
\end{equation}
where $\gamma$ is an arbitrary integration constant.
Substituting (\ref{eq:dilgamma}) in (\ref{eq:EOMa=0}), the last two
equations reduce to a single one, given by:
\begin{equation}
\label{eq:sqrt}
 \dot{a} = \epsilon\  \sqrt{\frac{\gamma^2}{9}\, a^{-6}
          -\frac{\Lambda}{12}\, a^{2} + \frac{\tilde{\Lambda}}{3}}\, ,
\end{equation}
where $\epsilon=\pm 1$ determines the branch of the square root chosen in the 
solution. The sign of the argument of the square root in (\ref{eq:sqrt}) will 
depend on the values of $\Lambda$ and $\tilde{\Lambda}$.
Both $\Lambda$ and $\tilde{\Lambda}$ are arbitrary and independent.
The solution only makes sense when this argument is positive. 
Integrating (\ref{eq:sqrt}) we obtain:
\begin{equation}
\label{eq:int}
 \int^{a}{\frac{\epsilon\, da'}{\sqrt{\frac{\gamma^2}{9}\, {a'}^{-6}   
   -\frac{\Lambda}{12}\, {a'}^{2} + \frac{\tilde{\Lambda}}{3}}}} = y+y_{0}\, .
\end{equation}
The l.h.s. of (\ref{eq:int}) cannot be expressed in terms of any elementary 
function, so it is hard to see whether a self-tuning mechanism could work.
As in \cite{KSS2}, it could be possible to study the bounds on the brane 
cosmological constant 
depending on the signs and values of $\Lambda$ and $\tilde{\Lambda}$.
We leave this question open for future research.

\vspace{.8cm}

\noindent
{\bf Acknowledgments}\\

\noindent
We would like to thank C\'esar G\'omez, Patrick Meessen and Tom\'as 
\Ortin for useful discussions. We also thank Jim Cline and Christophe 
Grojean for pointing out some problems in a previous version.
The work of N.A.A. and B.J. has been 
supported by the TMR program FMRX-CT96-0012 on {\sl Integrability, 
non-perturbative effects, and symmetry in quantum field theory}. 


\end{document}